# Computational paper wrapping transforms non-stretchable 2D devices into wearable and conformable 3D devices

Yu-Ki Lee[1†], Zhonghua Xi[2†], Young-Joo Lee[1], Yun-Hyeong Kim[1], Yue Hao[2], Young-Chang Joo[1], Changsoon Kim[3], Jyh-Ming Lien[2*], In-Suk Choi[1*]

[1]Department of Materials Science and Engineering, Seoul National University, Seoul 08826, Republic of Korea.

[2]Department of Computer Science, George Mason University, Fairfax, VA 22030, USA.

[3]Graduate School of Convergence Science and Technology, and Inter-University Semiconductor Research Center, Seoul National University, Seoul 08826, Republic of Korea.

*Correspondence to: jmlien@cs.gmu.edu and insukchoi@snu.ac.kr

**Abstract:** This study starts from the counter-intuitive question of how we can render a conventional stiff, non-stretchable and even brittle material conformable so that it can fully wrap around a curved surface, such as a sphere, without failure. Here, we answer this conundrum by extending geometrical design in computational *kirigami* (paper cutting and folding) to paper wrapping. Our computational paper wrapping-based approach provides the more robust and reliable fabrication of conformal devices than paper folding approaches. This in turn leads to a significant increase in the applicability of computational *kirigami* to real-world fabrication. This new computer-aided design transforms 2D-based conventional materials, such as Si and copper, into a variety of targeted conformal structures that can fully wrap the desired 3D structure without plastic deformation or fracture. We further demonstrated that our novel approach enables a pluripotent design platform to transform conventional non-stretchable 2D-based devices, such as electroluminescent lighting and a paper battery, into wearable and conformable 3D curved devices.

**One Sentence Summary:** Computational paper wrapping provides a novel way to wrap an arbitrary curved 3D shapes with 2D based brittle materials, such as Si and Copper.

**Main Text:** New flexible and stretchable materials have recently attracted significant interest for fabricating conformable devices(*1-4*). However, their application for real devices remains strictly limited by their failure to provide certain key criteria available in conventional devices, such as high conductivity and reliability. This study introduces a novel way to utilize conventional non-stretchable materials to reliably wrap any arbitrary 3D curved surfaces, including the human body and curved vehicle interiors as potential applications. This new approach allows us to make conformable devices without sacrificing their performance.

Wrapping an arbitrary surface tightly with non-stretchable material is challenging(*5, 6*). As shown in Fig. 1A, wrapping a sphere with rectangular paper cannot avoid the formation of wrinkles and overlaps. If the material is a substrate or other active layer of a flexible device, severe deformation and overlapping in the material can cause fracture or breakage of the device as shown in Fig. S1. One common way to wrap 3D surfaces with non-stretchable materials is by cutting the materials, called *kirigami* (paper cutting and folding) design. A recent example is *kirigami*-patterned cuts, such as lattice patterns and fractal cut patterns. Yigil Cho et al. proposed a mechanical meta-material by cutting planar materials with a hierarchy auxetic structure called fractal cut(*7*). They showed that the fractal cut-designed materials are shape programmable and can effectively cover a sphere. In addition, Mina Konakovi'c et al. suggested a method for designing complex 3D models using 2D auxetic structures through computer algorithms(*8*). Both studies overcome the limitations of existing flexible devices by geometrically designing cut patterns. However, when covering 3D complex surfaces, these structures inevitably result in openings that lead to design and functional limitations in conformal devices, such as sub-optimal coverage of printable batteries and undesirable holes in lighting and displays (Fig. 1B).

To overcome this problem, the ingenuity of this work is that we introduce a computational method that is inspired from the process of peeling off a tangerine skin, as illustrated in Fig. 1C-D, called polyhedral edge unfolding in order to make a 2D non-stretchable material platform to make conformal and wearable devices. Notably, the patches created by arbitrarily cutting a curved surface, as one may peel the skin that fully covers a tangerine (Fig. 1C), cannot be a solution to make an appropriate 2D unfolding structure because they are usually non-developable surfaces. A series of snapshots are given in Fig. 1D to illustrate the 2D unfolding of a sphere generated by our algorithmic *kirigami* that can wrap a steel ball. Theoretically, a curved surface can be characterized by the Gaussian curvature, which is the vector product of maximum and minimum principal curvatures at a point. Two-dimensional materials with zero Gaussian curvature at all points, such as a sheet of paper, is called a developable surface and cannot be transformed into a 3D non-developable surface with positive or negative Gaussian curvatures without tearing, stretching or compressing. This has been mathematically proven by 'Gauss Theorema Egregium', which states that "To move a surface onto another surface must match the Gaussian curvature of all corresponding points(*9*)" (Fig. S2). Instead, polyhedral edge unfolding(*10, 11*) flattens arbitrary surfaces into 2D structures by cutting its surface into developable patches with $C^0$ continuity, i.e., patches are connected without gaps and seams. The computer science community has recently made great effort to algorithmically determine surface cuts that segment a non-developable surface into developable surface patches, called polyhedral nets (or simply nets)(*12-14*). Beyond designing valid nets, recent computational methods, including those developed by the authors(*15-18*), have focused on optimizing net quality and foldability using machine learning techniques that drastically reduce the time and effort needed in traditional trial-and-error approaches.

A polyhedron can have multiple possible nets(*19*), especially when it is convex. For a given polyhedral sphere, we demonstrated six unfoldings created by six different methods (Fig. S3), and Fig. 1F shows one of them, called "Steepest Edge (SE) Unfolding"(*12*). Given a polyhedron $P$ and a 3D vector $\vec{c}$, SE Unfolding associates each edge $e = (v, w)$ of $P$ with a weight $\frac{<c, v-w>}{|v-w|}$, where $v$ and $w$ are the end points of $e$ and $<..>$ denotes the inner product between two vectors. Then, for each vertex $v$, the edge incident to $v$ with the largest weight, i.e., most aligned with $\vec{c}$, is cut. SE Unfolding can unfold almost all convex polyhedra with a few randomly selected $\vec{c}$, creating radially spread unfolded figures, as in Fig. 1F. Other methods are also available, such as Flat-Tree Unfolding(*12*) (Fig. 1G), which finds the minimum spanning tree using the same weight as in SE Unfolding, and Genetic Algorithms (GA)(*18, 20*), to unfold non-convex polyhedra into the fewest possible patches (Fig. S3). As it is critically important that a conformal device is composed of one or as few pieces as possible, our recent GA methods can evolve the unfoldings by mutating the edge weights until a net with zero overlaps is found(*18*). The evolution is controlled by a fitness function $f : N \rightarrow \mathbb{R}$ that evaluates an unfolding. An example of $f(N)$ is defined as:

$$f(N) = -(\lambda_0 \delta_0 + \lambda_1 \delta_1),$$

where $\delta_0$ is the number of overlaps in N, $\delta_1$ is the number of hyperbolic vertices that cause local overlaps in $N$, and $\lambda_0$ and $\lambda_1$ are user parameters set to 100 and 1, respectively. Clearly, the maximum value of $f(N)$ must be 0, which yields a valid net.

These conventional computational methods, however, are mostly focused on and demonstrated with hollow 3D *kirigami* structures(*11, 13, 17, 21-24*) through creasing and folding, which do not have real application because of labor intensive and structurally weak fabrication process. By contrast, we adopt computational *kirigami* methods to cover an arbitrary curved surface, as exemplified in Fig. 1D, which can be applicable to make a conformal and wearable device. For instance, covering the complex General Lee Sun Shin statue (of 3000 facets) by attaching was significantly faster and structurally robust than making the hollow 3D statue by gluing individual edges, indicating that our method has advantages in handling complex shapes (Fig. S4). However, we still have the issue in folding the creases to cover the surface by conventional computational *kirigami* approach. As illustrated in Fig. 2A, we fabricated four different unfolding paper structures from four polyhedral spheres with different resolutions by computational *kirigami*. The surface area with 80 facets is 6.4% larger than that of a perfect sphere ($4\pi r^2$). The area difference is reduced as the mesh numbers increase, resulting in 1.1% for 500 facets. Consequently, increasing the resolution of meshes provides better conformability on the steel ball, as shown in Fig. 2B. In many applications of conformal devices, accurate covering and tight wrapping is crucial. Additionally, because most real-world objects are smooth and curved, high-resolution meshes are needed to provide the necessary covering and wrapping accuracy, but unfortunately, they require a long folding time. Here, we develop a new approach beyond conventional computational *kirigami*. By considering conformal device design as the paper wrapping problem instead of the paper folding (origami) problem, we recognize that the special operations of attaching and wrapping conformal devices to cover an underlying curved 3D surface are much easier by bending and pressing than creasing and folding. Paradoxically, we found that, with bending and attaching, high-resolution meshes provide us with the needed solution to address the limitations of long fabrication times. A mesh with high-resolution tessellation can better represent surfaces with a higher degree of continuity (e.g., $G^1$ continuity along the branches where facets on both sides of a crease line share a common tangent direction), which is common for many surfaces formed by real materials. Computationally, this can be achieved by modifying the GA fitness function to:
$$f(N) = \min(-(\lambda_0 \delta_0 + \lambda_1 \delta_1), \sum_{e_i} \theta_{e_i}),$$
where $\theta_{e_i}$ is the dihedral angle of a cut edge $e_i$ in the net *N*. This enhanced $f(N)$ further evolves *N* so that the sum of its folding angles is minimized, and thus optimized for bending, by maximizing $f(N)$. Therefore, as shown in Fig. 3A, given a polyhedra net of a high-resolution mesh, though we still use the outer cut lines of the net, we can ignore (erase) all the creases and simply bend the stripes of the net and tightly wrap the sphere without gaps, overlaps, or wrinkles. Therefore, more complex and smoother 3D surfaces can be fabricated accurately and approximately much times faster for handling complex shapes. Moreover, we note that the small bending angles of the stripes prevent the materials used in the conformal devices from being damaged when wrapping the object. These properties enable the application of our method to non-stretchable and stiff metallic materials. As shown in Fig. 3B, for the same computational paper wrapping pattern with 500 facets, Cu foil can tightly wrap a steel ball, similar to the paper shown in Fig. 3C. (Please see the supporting information, Fig. S5, which illustrates the case of paper wrapping pattern of 80 meshes with Cu foil resulting in wrinkles and uncovered area.) Moreover, we can even apply our method to brittle Si wafers. We fabricated a computational paper wrapping pattern to cover part of a sphere with a diameter of 4 cm, as shown in Fig. 3D-G. A thin Si wafer with a thickness of 20 µm was cut by laser cutting based on the computed unfolding structure consisting of 100 meshes without crease lines. We attached it on both a desired convex and a concave curved surface without gaps, overlaps, or fracture.

The above results have significant practical implications. Most devices are made from a 2D brittle substrate material that can nevertheless be slightly bent at a small angle if the aspect ratio is large enough(*25*). As long as we can make 2D unfolding structures, we can use conventional materials and processes, such as deposition, evaporation, etc., to wrap a curved surface with a conformable device. These in turn lead to a significant increase in the applicability of computational paper wrapping to real industrial fabrication processes by employing wrapping instead of folding in the fabrication of conformal devices. For instance, here, we realize a conformal device using commercial electroluminescent lamp (EL) panels. We simply cut the EL panels with computationally programmed wrapping pattern of a sphere with 500 meshes without crease lines using a laser cutter, including power supply terminals at the end of an edge site. The final cut EL panel can be stably and easily attached on the sphere (Fig. 4A-B). EL lighting consists of top and bottom brittle electrodes with solid interlayers that can be easily fractured by creasing and folding. However, there was no catastrophic failure, and the device demonstrated good operation, as shown in Fig. 4C, because we conformally wrapped the sphere by bending and pressing. In addition to a sphere model, a Korean traditional mask was conformably wrapped by EL panels. We used 3D scanning to obtain the 3D model and generate meshes for the most curved region. For the Korean traditional mask model, total of 169 meshes are used for the target region and are unfolded with the SE Unfolding method without overlapping (Fig. 4D and Fig. S6). Finally, the EL panel with computational wrapping pattern is conformally attached on the mask, which demonstrated good operation without failure (Fig. 4E-F). The concept was applied for the more complex exterior parts of an electrical toy vehicle having non-zero Gaussian surface. We generated the meshes and the corresponding paper wrapping patterns are generated with the GA unfolding method as shown in the Fig. S7 and then EL panels were cut and attached on the surfaces to decorate the car exteriors (Fig. 4G-I). We also fabricated a conformal battery and demonstrated the advantage of the concept with a complex 3D shape. For battery fabrication, the diversity of computational paper wrapping patterns can bring advantages for efficient manufacturing processes and reducing the waste of printing materials for screen prints (see supporting information, Fig. S8).

In summary, we introduce the concept of computational paper wrapping to transform non-stretchable 2D devices into 3D conformal devices. By extending the algorithmic *kirigami* method that cuts a polyhedral mesh into the fewest number of developable patches, we can prepare a conformable paper wrapping design of a 2D sheet for any 3D surfaces. Particularly, we have successfully shown that slightly bending non-stretchable substrates to comply with the underlying object can effectively wrap the desired shape without the formation of inherent crease lines between adjacent facets. Consequently, even brittle materials, such as metal films or Si wafers, can fully cover and tightly wrap non-zero Gaussian surfaces without catastrophic fracture. We demonstrated that the advantages of wrapping over folding can be applied to make real conformal devices from conventional devices, such as EL lighting devices and batteries. Our results provide new insights into the development of conformal devices with arbitrary shapes using efficient algorithms with robust and reliable conventional fabrication processes.

**References and Notes:**

1. C. Yang, Z. Suo, Hydrogel ionotronics. Nature Reviews Materials 3, 125-142 (2018).
2. H.-H. L. Chong-Chan Kim, 1* Kyu Hwan Oh,1,2 Jeong-Yun Sun1,2†, Highly stretchable, transparent ionic touch panel. Science 353, 682-687 (2016).
3. S. Lin et al., Stretchable Hydrogel Electronics and Devices. Adv Mater 28, 4497-4505 (2016).
4. J. A. Rogers, Takao Someya, and Yonggang Huang, Materials and mechanics for stretchable electronics. science 327.5973, 1603-1607 (2010).
5. J. Stillwell, Mathematics and Its History. Springer Science & Business Media, 343-348 (2010).
6. J. Hure, B. Roman, J. Bico, Wrapping an adhesive sphere with an elastic sheet. Phys Rev Lett 106, 174301 (2011).
7. Y. Cho et al., Engineering the shape and structure of materials by fractal cut. Proc Natl Acad Sci U S A 111, 17390-17395 (2014).
8. M. Konaković et al., Beyond developable. ACM Transactions on Graphics 35, 1-11 (2016).
9. M. P. Do Carmo, Differential Geometry of Curves and Surfaces: Revised and Updated Second Edition. . Courier Dover Publications,  (20016).
10. J. S. McCranie, An investigation of the size of epsilon-nets. Univ. Illinois, Urbana, IL,  (1986).
11. E. D. D. a. J. O'Rourke, Geometric folding algorithms. Cambridge university press Cambridge,  (2007).
12. W. Schlickenrieder, Nets of polyhedra. Citeseer,  (1997).
13. V. K. D. Julius, and A. Sheffer, D-Charts: Quasi-Developable Mesh Segmentation. Computer Graphics Forum 24, 581-590 (2005).
14. E. D. M. Bern, D. Eppstein, E. Kuo, A. Mantler, and J. Snoeyink, Ununfoldable Polyhedra with Convex Faces. Comput. Geom. Theory Appl. 24, 51-62 (2003).
15. Y.-H. K. Y. Hao, Z. Xi, and J.-M. Lien, Creating Foldable Polyhedral Nets Using Evolution Control. in Proceedings of the Robotics: Science and Systems Conference (RSS),  (2018).
16. Y.-H. K. Y. Hao, and J.-M. Lien, Synthesis of Fast and Collision-free Folding of Polyhedral Nets. in Proceedings of the ACM Symposium on Computational Fabrication (SCF),  (2018).
17. Z. X. Y.-H. Kim, and J.-M. Lien, Disjoint convex shell and its applications in mesh unfolding. CAD Comput. Aided Des. 90,  (2017).
18. Z. Xi, Y.-h. Kim, Y. J. Kim, J.-M. Lien, Learning to segment and unfold polyhedral mesh from failures. Computers & Graphics 58, 139-149 (2016).
19. P. F. D. P. M. Dodd, and S. C. Glotzer, Universal folding pathways of polyhedron nets. Proc. Natl. Acad. Sci.,  (2018).
20. H.-Y. W. S. Takahashi, S. H. Saw, C.-C. Lin, and H.-C. Yen, Optimized topological surgery for unfolding 3d meshes. Computer Graphics Forum 30, 2077-2086 (2011).
21. T. Tachi, Origamizing polyhedral surfaces. Vis. Comput. Graph. IEEE Trans. 16, 298-311 (2010).
22. J. Mitani, Hiromasa Suzuki, Making Papercraft Toys from Meshes using Strip-based Approximate Unfolding. ACM transactions on graphics (TOG) 23, 259-263 (2004).

23. A. T. I. Shatz, and G. Leifman, Paper craft models from meshes. Vis. Comput. 22, 825-834 (2006).
24. G. C. Massarwi F, Elber G. , Papercraft models using generalized cylinders. In: 15th Pacific conference on computer graphics and applications PG157–148 ,07′ (2007).
25. M. A. Mahmoud, and A. Hosseini, Assessment of stress intensity factor and aspect ratio variability of surface cracks in bending plates. Engineering fracture mechanics 24, 207-221 (1986).

**Acknowledgments:** This work was supported by NSF IIS-096053, CNS-1205260, EFRI-1240459, AFOSRFA9550-12-1-0238, Creative-Pioneering Researchers Program through Seoul National University, and Research Resettlement Fund for the new faculty of Seoul National University

**Supplementary Materials:**

Materials and Methods

Figs. S1-S8

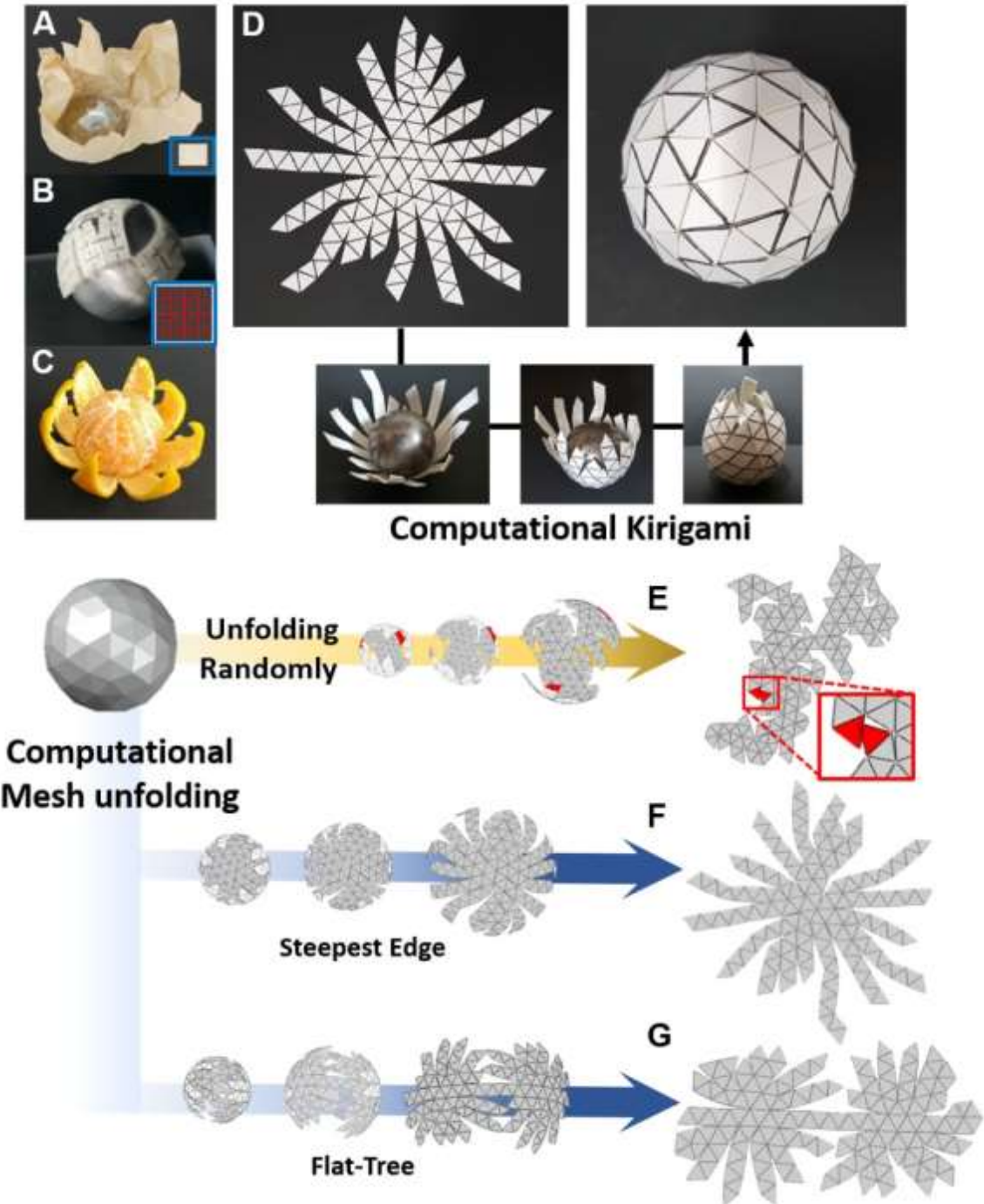

**Fig. 1. Reverse engineering computational *kirigami* for conformable wrapping.**
(**A**) Wrinkles are formed when wrapping a rectangular sheet of paper tightly around a non-zero Gaussian surface. (**B**). Cut-pattern-*kirigami* can avoid wrinkles but leads to inevitable openings and uncovered areas. (**C-D**). Peeling a tangerine's skin inspired our algorithmic *kirigami* design. (**E**) However, arbitrary cuts (peeling) cannot be guaranteed to create flat and non-overlapping 2D structures. Computational mesh unfolding studies how to algorithmically cut and unfold the meshed shapes without overlapping, including methods such as (**F**) Steepest Edge Unfolding and (**G**) Flat-Tree Unfolding.

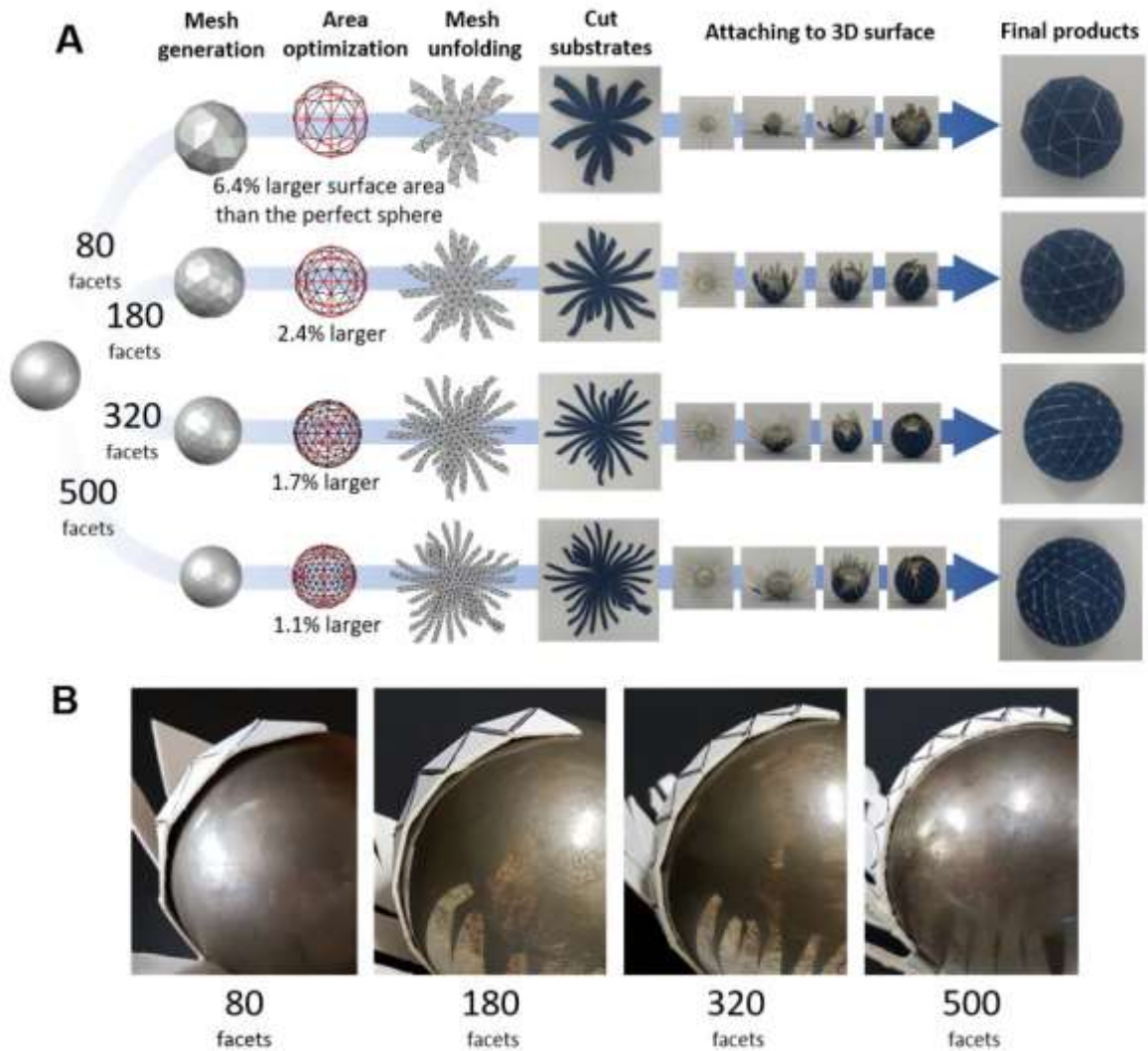

**Fig. 2. Demonstration with thick, rigid card stock.** (**A**) As the number of facets increases, the smoothness and conformability of the mesh improve naturally. The difference in surface area between the perfect sphere and the approximated unfolded figures reduces 5.3% when going from 80 facets to 500 facets. (**B**) Naturally, increasing the resolution of meshes provides better conformability on the steel ball.

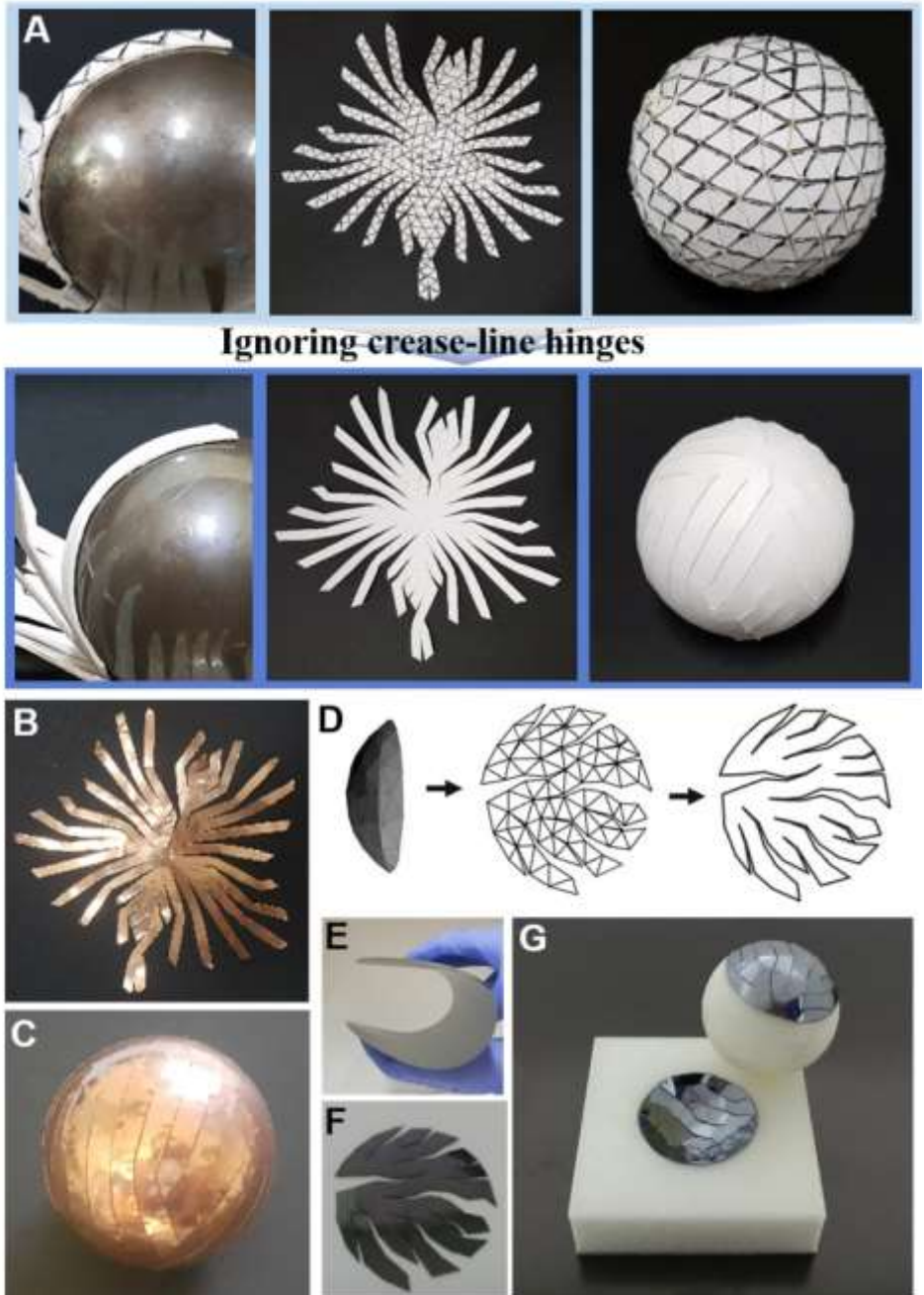

**Fig. 3. Ignoring crease-line hinges for better applicability to diverse materials.** When the total folding angle of a net is minimized, crease lines can be ignored to accommodate flexible but non-stretchable, stiff materials. (**A**) For 500 facets, however, the gaps in the case of a rigid material and the wrinkles in the case of a flexible material are reduced so as to be invisible, and the difference between the two becomes imperceptible. (**B**) Non-stretchable Cu foil is cut into the computational paper wrapping pattern. (**C**) With a sufficient number of facets, Cu foil can be bent and fully wrap a sphere without creasing or folding. (**D**) Part of the sphere is unfolded with 100 meshes, and the crease lines are removed. (**E-F**) A 20 µm-thick brittle Si wafer is cut into the unfolded net by a laser cutter, and (**G**) the cut Si wafer can wrap both convex and concave frames stably.

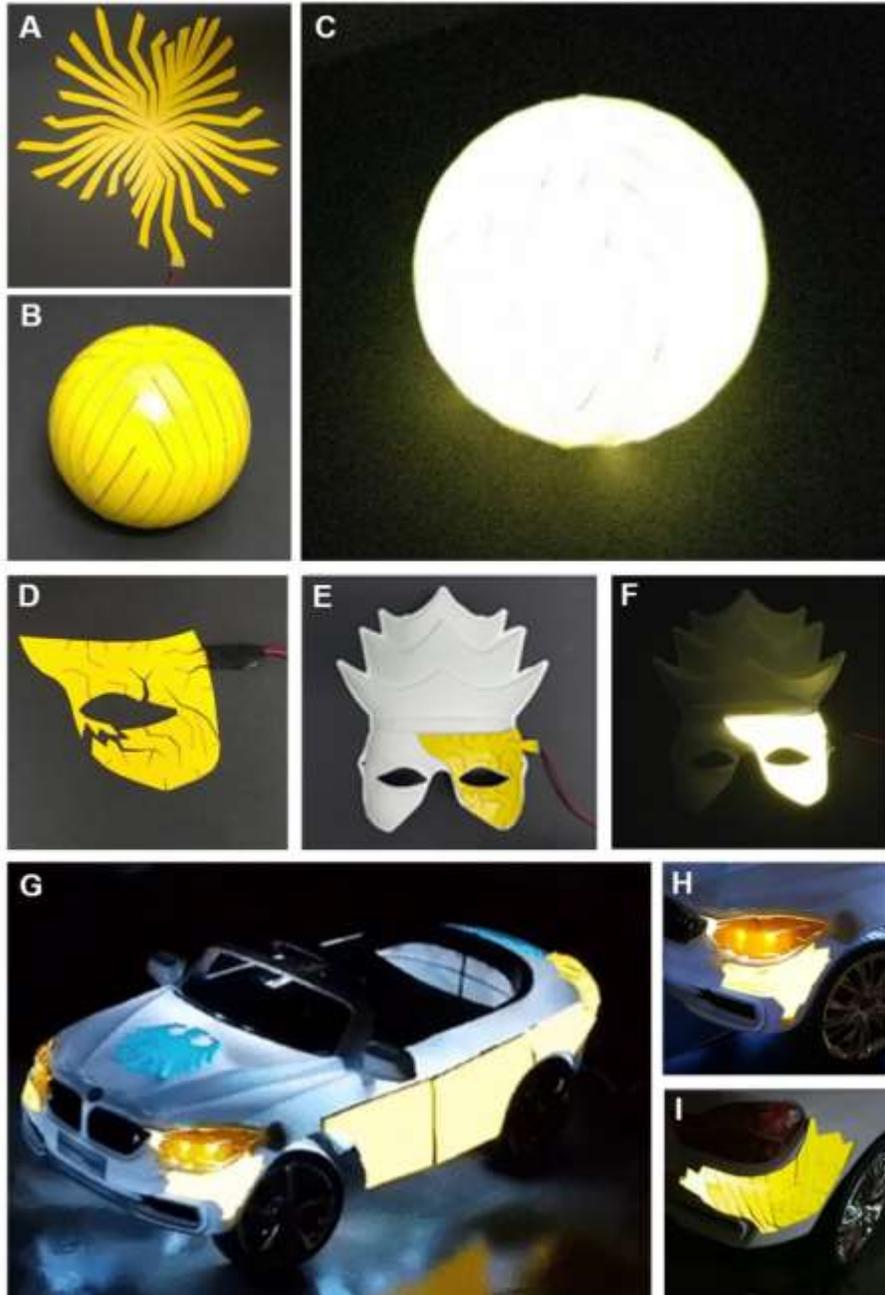

**Fig. 4. Demonstration of conformable devices.** (**A**) Cuttable, non-stretchable EL lighting consisting of brittle electrodes is cut with a laser cutter in the computational paper wrapping pattern for a spherical mesh with 500 facets. (**B**) EL lighting with the computational paper wrapping pattern can fully cover a sphere and (**C**) operate without catastrophic failure. (**D-F**) In addition to a sphere model, practically used facial mask can also be conformably covered with EL lighting and operated without electrical failure. (**G**) An electrical toy vehicle also can be conformably wrapped with EL panels in the same way, and the attached EL panels are well operated without failure as well. The GA unfolding method is used for generating computational paper wrapping patterns for parts having non-zero Gaussian surface, such as (**H**) the headlights, the edge of the front side bumper, and (**I**) the edge of the rear side bumper of the electrical toy vehicle.

# Supplementary Materials for

## Computational paper wrapping transforms non-stretchable 2D devices into conformable 3D devices

Yu-Ki Lee[1†], Zhonghua Xi[2†], Young-Joo Lee[1], Yun-Hyeong Kim[1], Yue Hao[2],

Young-Chang Joo[1], Changsoon Kim[3], Jyh-Ming Lien[2*], In-Suk Choi[1*]

Correspondence to: jmlien@cs.gmu.edu and insukchoi@snu.ac.kr

**This PDF file includes:**

    Materials and Methods
    Figs. S1 to S8

**Materials and Methods**

We implemented the proposed stacking algorithm in C++. All data are collected on a MacBook Pro with a 2.5 GHz Intel Core i7 CPU with 16 GB Memory running macOS 10.12. For experimental demonstration, we used a 1 mm-thick polyurethane sheet, a 125 µm-thick polyimide film, a 30 µm-thick copper foil, a 20 µm-thick Si wafer, and commercial EL lighting panels.

**Supplementary Figures**

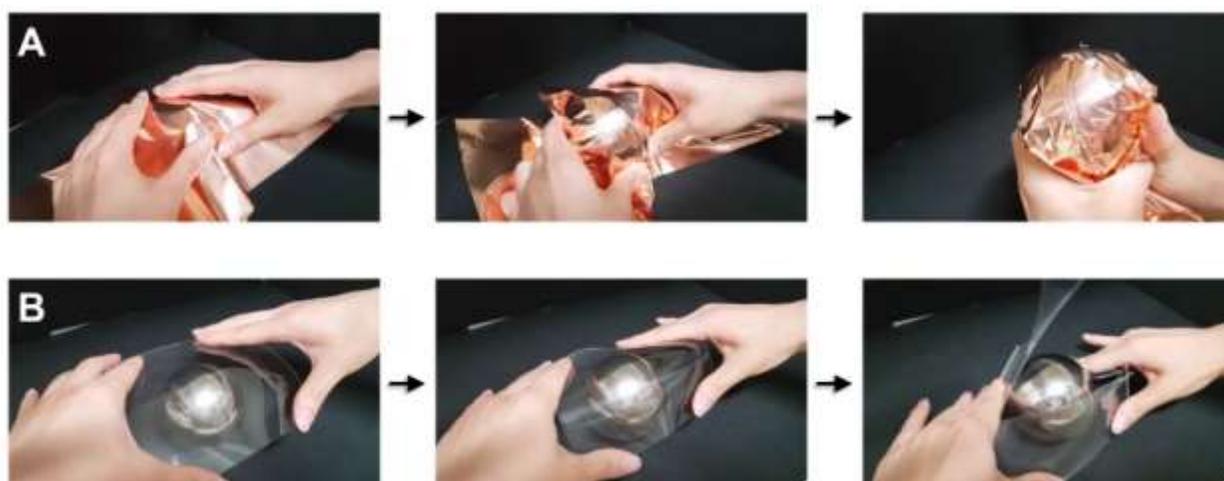

**Fig. S1.** Failure of covering the sphere by non-stretchable sheet, such as **(A)** a 30 µm-thick Cu foil and **(B)** a glass substrate of commercial LCD panel.

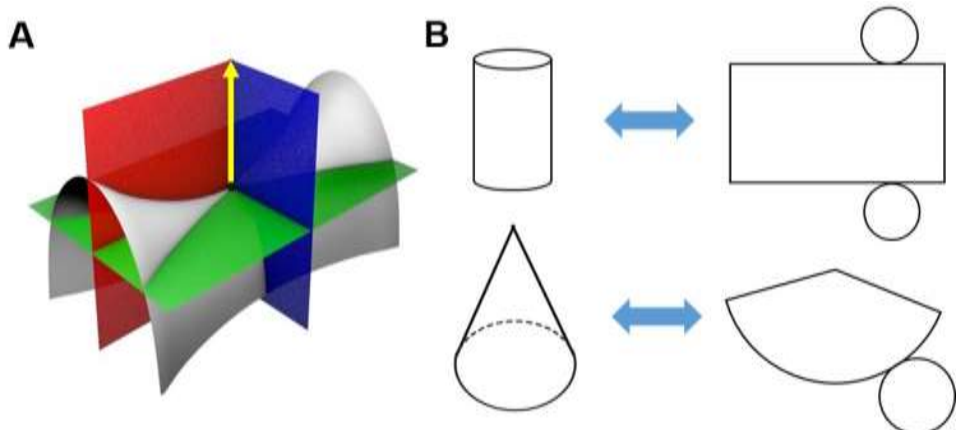

**Fig. S2. (A)** Gaussian curvature is the vector product of the maximum and minimum values of curvature at a point, called the principal curvatures. At the saddle point (black) of the gray surface, one of the principal curvatures is the intersection between the red and gray surfaces, and the other is the intersection between the blue and gray surfaces. Both the red and blue planes contain the normal vector of the saddle point, and their intersections with the gray surface define the principal curvatures. A 2D material having zero Gaussian curvature points, such as a sheet of paper, called a "developable surface", cannot be transformed into a 3D surface with a positive or negative Gaussian curvature, called a "non-developable surface", without stretching or compressing. **(B)** For example, a cylinder or a cone can be covered with cut paper, but a saddle or a sphere cannot be wrapped without the formation of wrinkles or cuts. The reverse (flattening) process is also the same, which is why there are distortions in the planar map of the Earth.

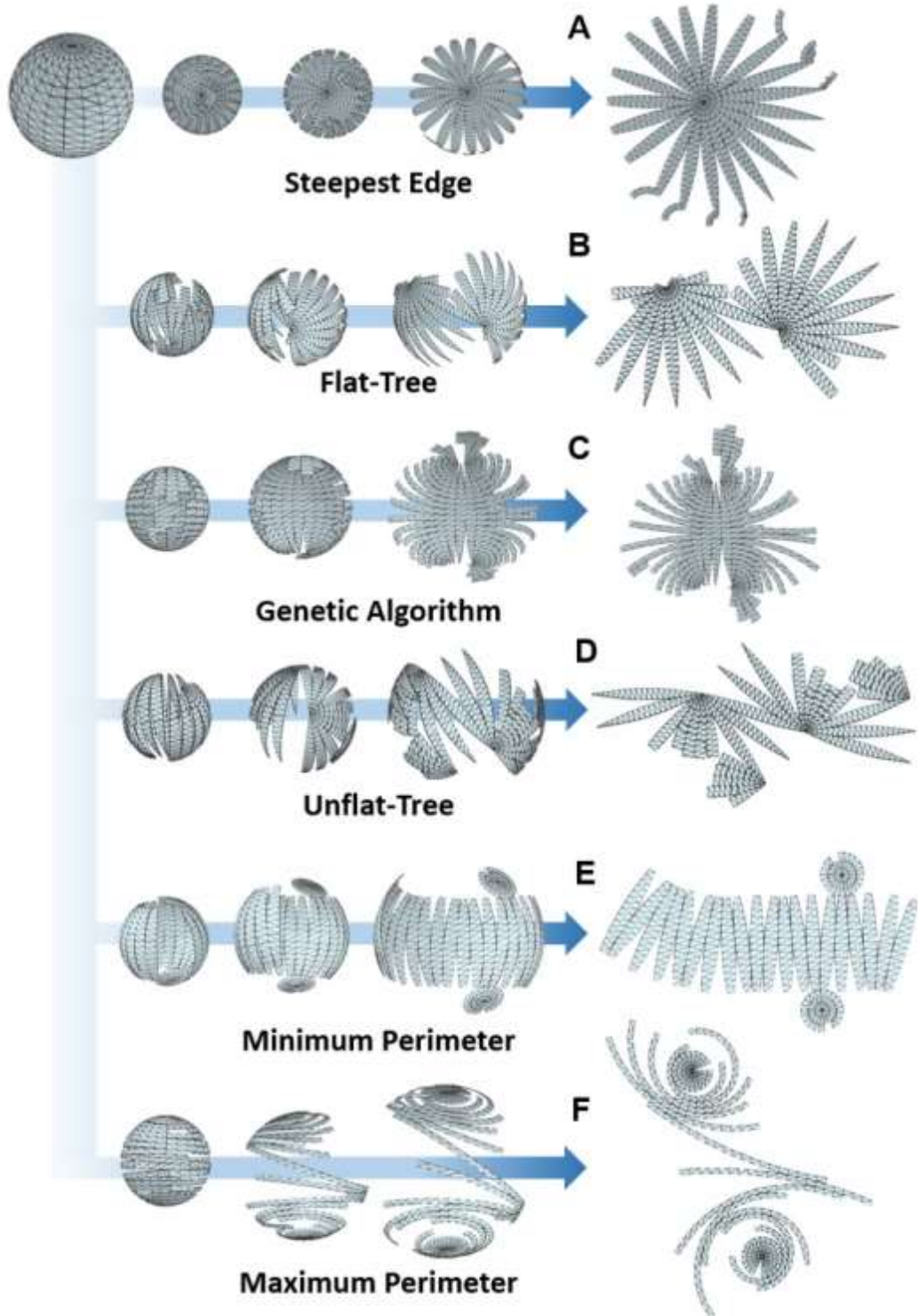

**Fig. S3. Diverse mesh unfolding methods were developed.** (**A**) Steepest Edge Unfolding;
(**B**) Flat-Tree Unfolding; (**C**) unfolding obtained by the proposed Genetic Algorithm, optimized to reduce the cut length; (**D**) Unflat-Tree Unfolding, with an edge weight $w'_i$ equal to $w'_i = 1 - w_i$, where $w_i$ is the edge weight used in Flat-Tree Unfolding; (**E**) Minimum Perimeter; and
(**F**) Maximum Perimeter.

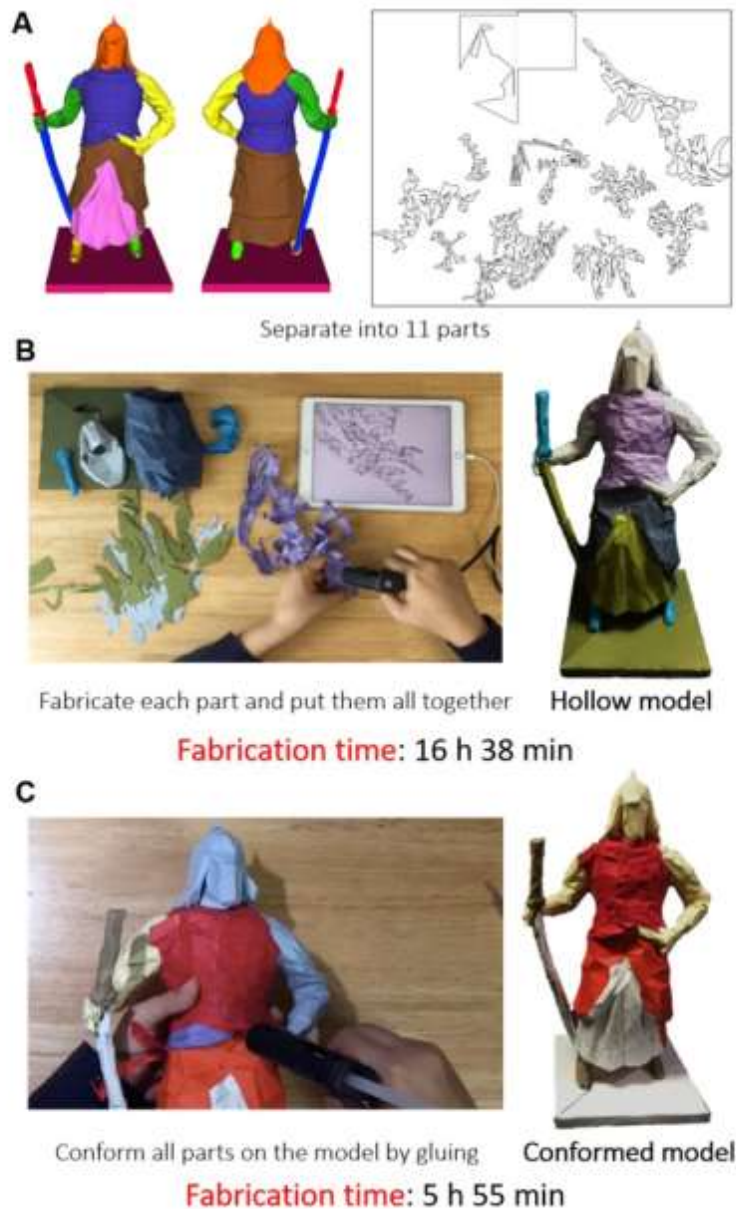

**Fig. S4. Practicality of covering the surfaces compared to folding the hollow 3D structure**. Either cutting a polyhedral mesh into nets or approximating with developable surfaces, segmentation of the mesh (the process of breaking a mesh into multiple components) can also be implemented by computational methods. Recent work from the authors made a breakthrough in segmentation by decoupling unfolding operations with a strategy that produces polyhedral nets by tightly coupling the edge unfolding and surface segmentation operations. Although most likely all of these initial unfolding structures will contain overlaps, the proposed method learns from these failures and identifies parts that may be unfolded into valid nets. **(A)** A mesh (3000 triangles) of a General Lee Sun Shin statue is unfolded into 11 patches, and the fabrication time between **(B)** making the hollow model and **(C)** wrapping on the 3D structure is compared. The process time of wrapping is almost three times shorter than making the hollow 3D structure, and it is mechanically sturdier and more stable. It should also be noted that wrapping without folding is even much easier than covering with folding.

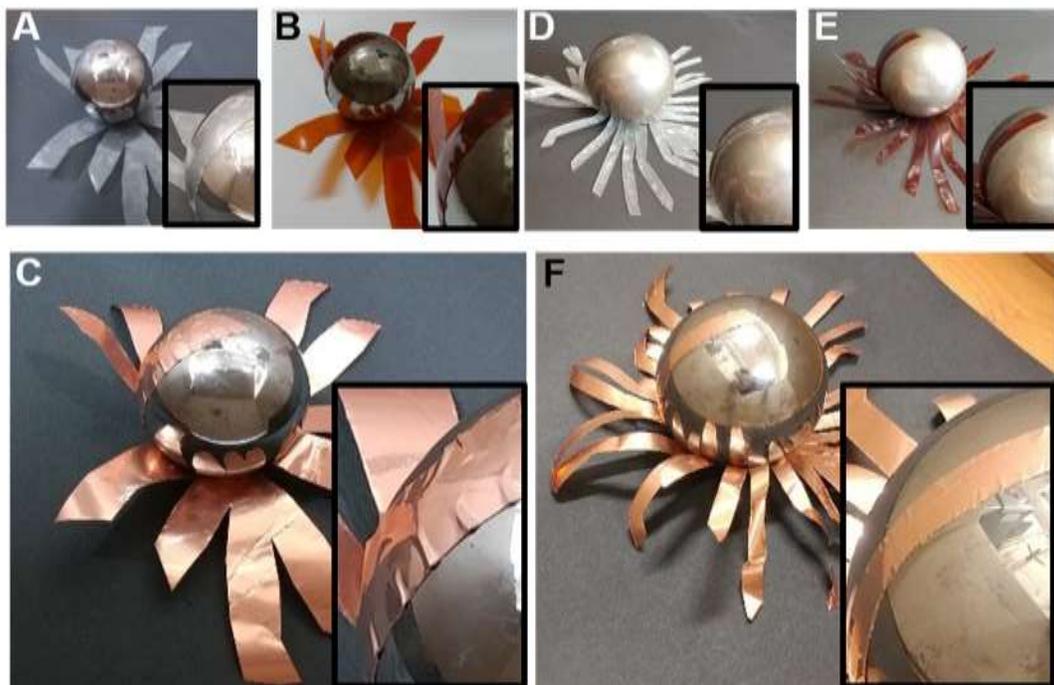

**Fig. S5.** A higher resolution of meshes is needed to conformally wrap a sphere with non-stretchable materials. With 80 meshes, **(A)** a polyurethane (PU) sheet can wrap the steel ball without wrinkles, and **(B)** a polyimide (PI) film has a few wrinkles only at the edges.
**(C)** However, Cu foil is severely deformed, and wrinkles are easily observed. On the other hand, by increasing the resolution to 500 meshes, conformable wrapping is possible with all three materials, and there are no distinguishable differences among the **(D)** PU sheet, **(E)** PI film, and **(F)** Cu foil.

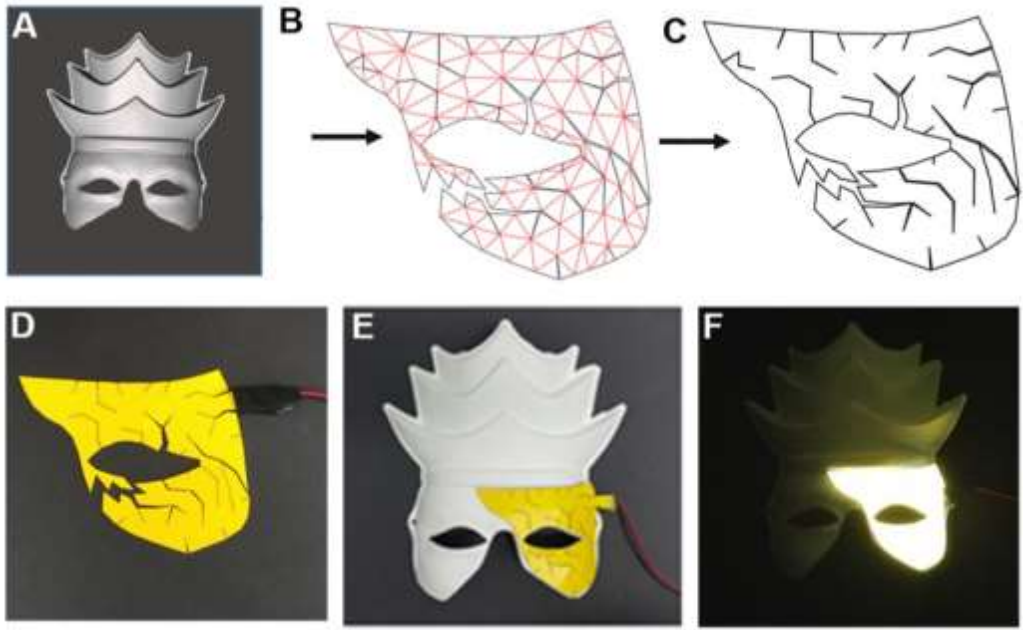

**Fig. S6.** Korean traditional masks can be covered with commercial EL panels. **(A)** The mask is 3D scanned, and a complex part containing both positive and negative Gaussian curvatures is re-meshed with 169 facets. **(B)** That part is then mesh unfolded, and **(C)** crease lines are ignored to apply our bending and attaching concept. **(D)** The commercial EL panel is cut by a laser cutter and **(E)** can be attached stably on the mask. **(F)** The light can be operated without failure, and the cut seams are blurred and invisible.

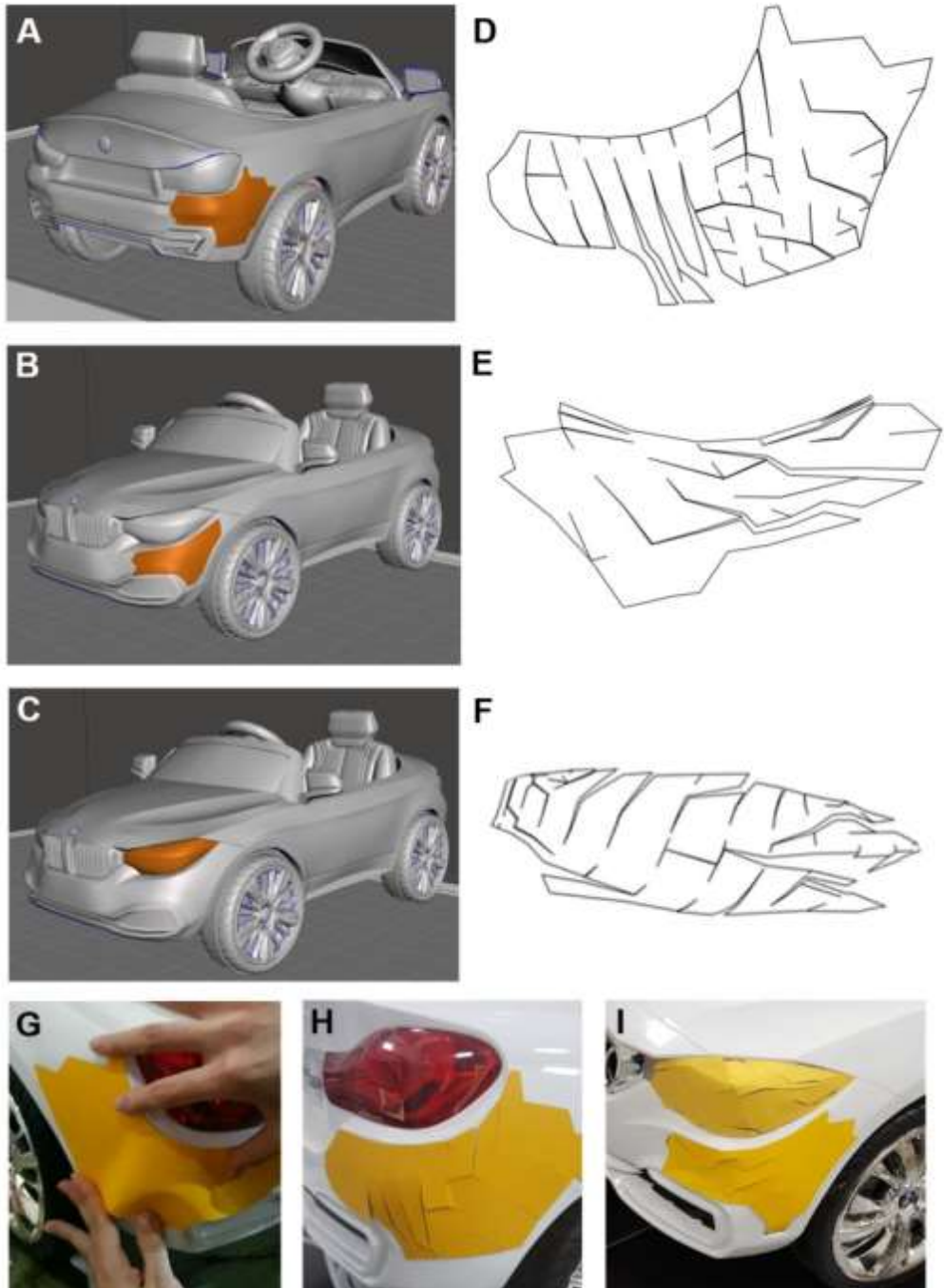

**Fig. S7.** The meshes were generated for the non-zero Gaussian surfaces in the electrical toy vehicle such as (**A**) the edge of the rear side bumper, (**B**) the edge of the front side bumper, and (**C**) the headlights, and then algorithmically unfolded by using the GA unfolding method respectively (**D-F**). Without having the computational paper wrapping pattern, (**G**) the rear side bumper region cannot be attached conformably and make gap between the simple cut pattern and surface of the electrical vehicle. However, the computational paper wrapping patterns (**H and I**) are conformably attached easily without gaps and overlapping, and the attached EL panels are well operated without any electrical failure.

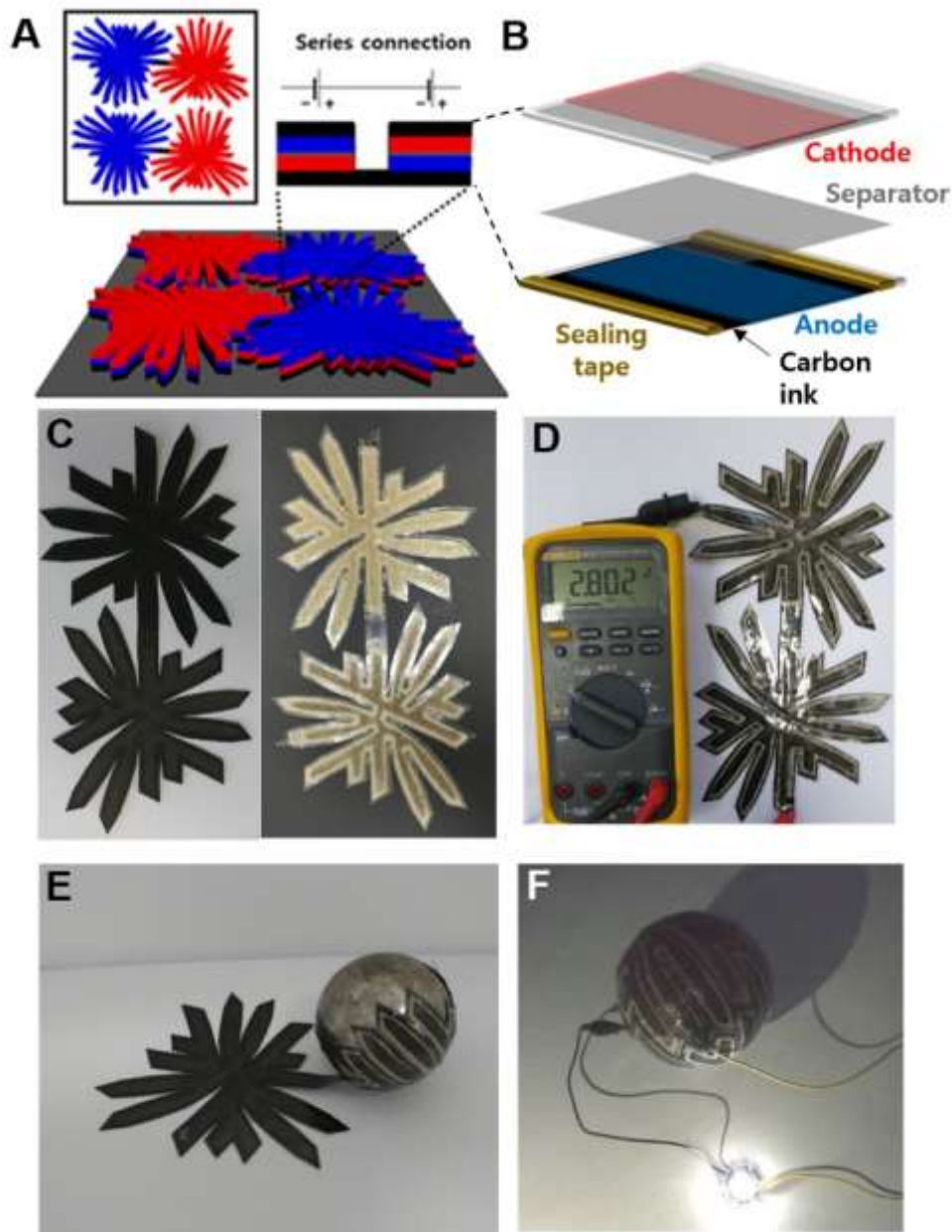

**Fig. S8.** Our computational paper wrapping concept is demonstrated with a screen-printed flexible Zn-carbon primary battery. (**A**) With the Flat-Tree Unfolding method, two unfolded figures can be located on one stencil, which indicates a more efficient manufacturing process and reduces waste of printing materials. (**B-C**) On a 50 μm-thick PET film with an unfolded figure shape, the carbon ink and Zn power-based slurry are printed and dried sequentially. For the other reflected-shaped PET film, the carbon ink and $MnO_2$ powder-based slurry are printed with the same method. Finally, by locating and sealing the cut separator wetted by the electrolyte between two PET films, we created a sphere-conformable battery. (**D**) In addition, serial connection of two cells can be achieved more easily with the Flat-Tree Unfolding method by making a "bridge" between two radial shapes. A series connection can be easily implemented, and the final Zn-carbon battery can have a 2.8 open circuit voltage. (**E**) It can wrap the steel ball stably and (**F**) can illuminate a LED lamp that requires more than 1.5 V to operate.